\documentclass{aa}  
\usepackage{graphicx}
\usepackage[varg]{txfonts}
\usepackage{hyperref}
\usepackage{natbib}
\usepackage{lscape}
\bibpunct{(}{)}{;}{a}{}{,}
\usepackage{epstopdf}
\begin{document}

\title{Abundances of disk and bulge giants from hi-res optical spectra\thanks{Based on observations collected at the European Southern Observatory, Chile (ESO programs 71.B-0617(A), 073.B-0074(A), and 085.B-0552(A))}}
\subtitle{II. O, Mg, Ca, and Ti in the bulge sample}
\author{H.~J\"onsson\inst{1,2,3} \and N.~Ryde\inst{1} \and M.~Schultheis\inst{4} \and M.~Zoccali\inst{5,6}}
\institute{Lund Observatory, Department of Astronomy and Theoretical Physics, Lund University, Box 43, SE-221 00 Lund, Sweden\\ \email{henrikj@astro.lu.se}\and
Instituto de Astrofísica de Canarias (IAC), E-38205 La Laguna, Tenerife, Spain\and 
Universidad de La Laguna, Dpto. Astrofísica, E-38206 La Laguna, Tenerife, Spain\and
Observatoire de la Cote d'Azur, Boulevard de l'Observatoire, B.P. 4229, F 06304 NICE Cedex 4, France\and 
Instituto de Astrof\'isica, Pontificia Universidad  Cat\'olica de Chile, Av. Vicu\~na Mackenna 4860, 782-0436 Macul, Santiago, Chile\and
Millennium Institute of Astrophysics, Av. Vicu\~na Mackenna 4860, 782-0436 Macul, Santiago, Chile
}
	    
\date{Submitted 2016; accepted 2016}

\abstract
   {Determining elemental abundances of bulge stars can, via chemical evolution modeling, help to understand the formation and evolution of the bulge. Recently there have been claims both for and against the bulge having a different [$\alpha$/Fe] vs. [Fe/H]-trend as compared to the local thick disk possibly meaning a faster, or at least different, formation time scale of the bulge as compared to the local thick disk.}
   {We aim to determine the abundances of oxygen, magnesium, calcium, and titanium in a sample of 46 bulge K-giants, 35 of which have been analyzed for oxygen and magnesium in previous works, and compare them to homogeneously determined elemental abundances of a local disk sample of 291 K-giants.} 
   {We use spectral synthesis to determine both the stellar parameters as well as the elemental abundances of the bulge stars analyzed here. The method is exactly the same as was used for analyzing the comparison sample of 291 local K-giants in Paper I of this series.}
   {Compared to the previous analysis of the 35 stars in our sample, we find lower [Mg/Fe] for [Fe/H]$>-0.5$, and therefore contradict the conclusion about a declining [O/Mg] for increasing [Fe/H]. We instead see a constant [O/Mg] over all the observed [Fe/H] in the bulge. Furthermore, we find no evidence for a different behavior of the alpha-iron trends in the bulge as compared to the local thick disk from our two samples.}
     {} 

\maketitle

% ##################### Introduction ##################
\section{Introduction} \label{sec:introduction}
The Galactic bulge holds a significant part of the stars of our Galaxy, but its history and evolution is still unknown. From cosmological $\Lambda$CDM-models it is expected that the bulge was formed via mergers of smaller dwarf galaxies, but recently it has been repeatedly shown that a major part of the bulge is dynamically formed from the inner disk, for example the fact that it shows cylindrical rotation \citep{2007ApJ...658L..29R,2012AJ....143...57K}, and that it has two red clumps \citep{2010ApJ...724.1491M}. On the other hand, the two red clumps might not be visible for lower-metallicity stars \citep{2012ApJ...756...22N,2013MNRAS.432.2092N}, and old RR Lyrae stars trace a component that is less elongated and is rotating slower \citep{2013ApJ...776L..19D,2016ApJ...821L..25K}, suggesting that there possibly is an old spheroid-bulge co-existing with the dynamically formed bar.

\citet{2012MNRAS.421..333S} predict that a classical merger-formed spheroidal bulge, if present, would have been spun up to bar-kinematics and therefore impossible/hard to find using kinematics alone, but the possible two stellar populations need to be distinguished by chemical differences. This is something that has been attempted several times during recent years: for example \citet{2007A&A...465..799L} measured O, Na, Mg, and Al in 53 bulge giants, 35 of which overlap with our sample. \citet{2010A&A...513A..35A} determined O, Na, Mg, Al, Si, Ca, and Ti in 25 bulge giants and when comparing to 55 similarly analyzed local giants (thin disk, thick disk, and halo), they find that the bulge has had a chemical evolution similar to the local thick disk. More recently \citet{2013A&A...549A.147B} used micro-lensing to observe 58 bulge dwarf and sub-giant stars, finding a wide age distribution with several young stars, a broad MDF, possibly including several components, and, when comparing to a similar sample of Solar neighborhood stars, they conclude that the `knee' in the [alpha/Fe] vs. [Fe/H] plots is shifted to $\sim0.1$ dex higher metallicities in the bulge, suggesting a faster chemical enrichment. Even more recently, \citet{2015A&A...584A..46G} analyzed slightly lower resolution spectra (the high-resolution GIBS-sample using FLAMES GIRAFFE with $R\sim 22500$) of 400 bulge giants in four bulge fields, finding a knee in their [Mg/Fe] vs [Fe/H] plot around [Fe/H]$\simeq-0.44$ dex, approximately 0.1 dex lower than \citet{2013A&A...549A.147B}. However, they lack a large enough, similarly observed and analyzed, solar neighborhood sample of giants to compare their bulge results to. They are therefore not able to conclude whether the difference in their sample compared to that of the microlensed dwarfs is due to systematic differences between the studies, or due to the very differently sized samples. \citet{2014AJ....148...67J} determine O, Na, Mg, Al, Si, Ca, Cr, Fe, Co, Ni, and Cu in 156 giants, using FLAMES GIRAFFE $R\sim22500$ spectra, in two bulge fields in common with ours (B3 and BL), finding a higher knee for the bulge giants as compared to literature samples of local dwarf stars \citep{2003A&A...410..527B,2005A&A...433..185B,2006MNRAS.367.1329R}. The fact that they are using two different types of stars, giants in the bulge and dwarfs in the local disk, analyzed differently, might introduce systematic differences that could account for the different position of the knee. In spite of the efforts put into these and many more works, still no consensus on the absolute abundance trends of the bulge and its evolution is reached. This is mainly because observing stars in the bulge is hard: it is situated relatively far away and covered behind dust in the disk. To handle these problems, one could do one or more of the following: go down in spectral resolution, thereby sacrificing abundance precision, possibly only enabling determination of the general metallicity (for example the low-resolution GIBS-sample, \citet{2014A&A...562A..66Z}), observe in the infrared, where determining the stellar parameters still is a problem (for example \citet{2012ApJ...746...59R}), use microlensing events, thereby not being able to select your targets and their positions \citep{2013A&A...549A.147B}, and/or use long integration times (as is done in for example \citet{2007A&A...465..799L} and this work).

This paper (hereafter, Paper II) is the second in a series determining abundances of bulge giants from optical high-resolution spectra ($R\sim47000$). Because of the long integration times needed for observing stars in the bulge at this high resolution not many such observations have been attempted. The spectra that were first used in \citet{2006A&A...457L...1Z} are therefore an unique dataset that has been analyzed in several subsequent articles: \citet{2007A&A...465..799L}, \citet{2010A&A...509A..20R}, \citet{2013A&A...559A...5B}, and \citet{2016A&A...586A...1V}. \citet{2010A&A...509A..20R} re-determine the stellar parameters as derived in the original article of \citet{2007A&A...465..799L}, for a small subset of the stars, and show that their all-spectroscopic approach in some cases is giving significantly different results, possibly influencing some of the abundance determinations and conclusions in \citet{2007A&A...465..799L}, \citet{2013A&A...559A...5B}, and \citet{2016A&A...586A...1V}. In order to eliminate systematic differences and ensure a homogeneous, differential comparison, we attempt to re-determine these stellar parameters, add eleven similarly observed spectra in a new field even closer to the Galactic center, and determine the alpha abundances oxygen, magnesium, calcium, and titanium. Thereby we will re-determine the oxygen abundances of 35 stars from \citet{2006A&A...457L...1Z} and magnesium abundances the same 35 stars from \citet{2007A&A...465..799L}, opening up for an interesting comparison between the results. In \citet{2016arXiv161105462J} (Paper I) we presented a similarly analyzed local disk sample of 291 similar giants, their stellar parameters and the abundances of oxygen, magnesium, calcium, and titanium. There we found that our stellar parameters were accurate and precise with a low dispersion compared to benchmark values based on fundamentally determined stellar parameters, such as effective temperatures from angular diameter measurements \citep{2003AJ....126.2502M}, and surface gravities from asteroseismic measurements \citep{2012A&A...543A.160T,2014ApJS..211....2H}. Furthermore, we found that the derived abundance trends show similar scatter as the trends of other Solar neighborhood works using dwarfs \citep{2014A&A...562A..71B}. In this paper we will determine the same abundances for a bulge sample of 46 giants of similar type as the previously published Solar neighborhood sample, enabling a differential comparison of abundances in the bulge and in the local disk.

% ##################### Observations ##################
\section{Observations} \label{sec:observations}
\begin{figure*}[htp]
\centering
\includegraphics[width=170mm]{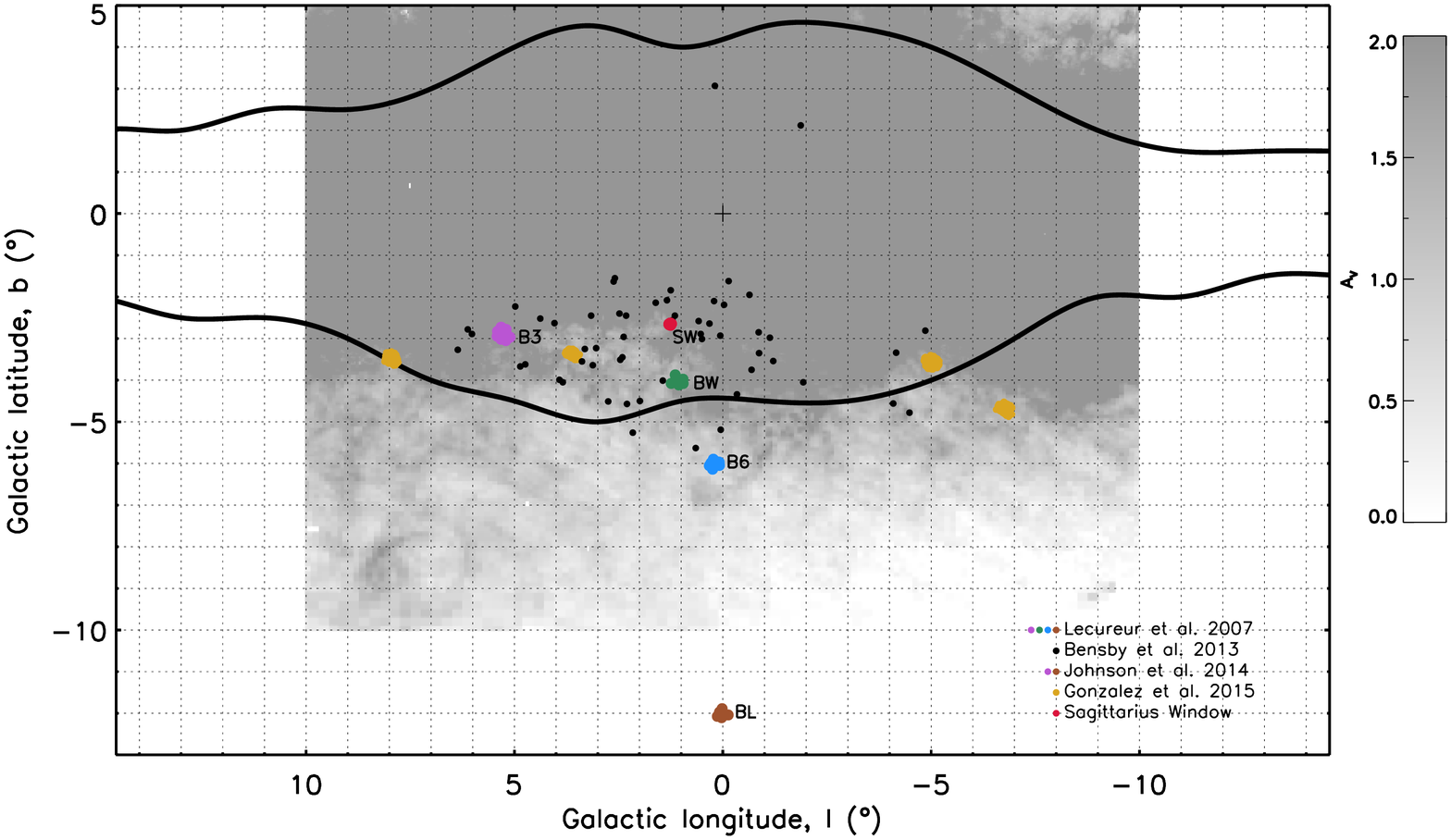}
\caption{Location of the five analyzed fields (B3, BW, B6, BL, and SW. All but the latter are previously analyzed in for example \citet{2007A&A...465..799L}) in comparison to the COBE/DIRBE outline of the Galactic bulge \citep{1994ApJ...425L..81W}, the locations of the microlensed bulge dwarf stars of \citet{2013A&A...549A.147B}, and the high-resolution sample of the GIBS survey \citep{2015A&A...584A..46G}. Also shown is the extinction towards the bulge from the BEAM calculator (http://mill.astro.puc.cl/BEAM/calculator.php) based on \citet{2011A&A...534A...3G,2012A&A...543A..13G} scaled to optical extinction \citep{1989ApJ...345..245C}.}
\label{fig:bulge_fields}
\end{figure*}

We have observed 46 K-giants in the Galactic bulge using the spectrometer FLAMES/UVES mounted on VLT.
The basic data of our stars are listed in Table \ref{tab:basicdata} and the Figure \ref{fig:bulge_fields} shows the location of our five fields (SW, B3, BW, B6, and BL) in comparison to the COBE/DIRBE outline of the Galactic bulge \citep{1994ApJ...425L..81W}, the locations of the microlensed bulge dwarfs of \citet{2013A&A...549A.147B}, and the high-resolution sample of the GIBS survey \citep{2015A&A...584A..46G}.
As can be seen in Figure \ref{fig:bulge_fields}, the new field in the Sagittarius Window (SW, $(l,b)=(1.25,-2.65)$) is closer to the Galactic centre than the other previously analyzed fields. Furthermore, it can be seen that it is situated in a region where the optical extinction is lower than the surroundings. To go even closer to the Galactic centre, infrared observations are needed due to the optical extinction being to high \citep[see e.g.][]{2015A&A...573A..14R,2016AJ....151....1R}; the corresponding infrared extinction in the bulge is is essentially zero outside of the plane (for $b<-1.5$ or $b>1.5$).

\begin{table}[htp]
\caption{Basic data for the observed bulge giants.}
\begin{tabular}{l c c c}
\hline
\hline
Star$^a$ & RA (J2000) & Dec (J2000) & $V$ \\
         & (h:m:s)    & (d:am:as)   &     \\
\hline
SW-09 &  17:59:04.533 & -29:10:36.53 & 16.153 \\
SW-15 &  17:59:04.753 & -29:12:14.77 & 16.326 \\
SW-17 &  17:59:08.138 & -29:11:20.10 & 16.388 \\
SW-18 &  17:59:06.455 & -29:10:30.53 & 16.410 \\
SW-27 &  17:59:04.457 & -29:10:20.67 & 16.484 \\
SW-28 &  17:59:07.005 & -29:13:11.35 & 16.485 \\
SW-33 &  17:59:03.331 & -29:10:25.60 & 16.549 \\
SW-34 &  17:58:54.418 & -29:11:19.82 & 16.559 \\
SW-43 &  17:59:04.059 & -29:13:30.26 & 16.606 \\
SW-71 &  17:58:58.257 & -29:12:56.97 & 16.892 \\
SW-76 &  17:58:54.192 & -29:12:09.31 & 16.943 \\
\hline
B3-b1 &  18:08:15.840 & -25:42:09.83 & 18.653 \\ 
B3-b5 &  18:09:00.527 & -25:48:06.78 & 18.345 \\
B3-b7 &  18:09:16.540 & -25:49:26.08 & 18.659 \\
B3-b8 &  18:08:24.602 & -25:48:44.39 & 18.915 \\
B3-f1 &  18:08:16.176 & -25:43:19.18 & 18.514 \\
B3-f2 &  18:09:15.609 & -25:57:32.75 & 18.924 \\
B3-f3 &  18:08:49.628 & -25:40:36.93 & 18.575 \\
B3-f4 &  18:08:44.293 & -26:00:25.05 & 18.650 \\
B3-f7 &  18:09:23.694 & -25:50:38.19 & 18.304 \\
B3-f8 &  18:08:12.632 & -25:50:04.45 & 18.490 \\
BW-b1 &  18:03:34.710 & -29:54:33.80 & 19.222 \\
BW-b2 &  18:04:23.950 & -30:05:57.80 & 18.484 \\
BW-b5 &  18:04:13.270 & -29:58:17.80 & 18.832 \\
BW-b6 &  18:03:51.840 & -30:06:27.90 & 18.410 \\
BW-b8 &  18:04:02.870 & -30:01:29.20 & 18.146 \\
BW-f1 &  18:03:37.140 & -29:54:22.30 & 17.994 \\
BW-f5 &  18:04:39.620 & -29:55:19.80 & 17.481 \\
BW-f6 &  18:03:36.890 & -30:07:04.30 & 18.387 \\
BW-f7 &  18:04:43.920 & -30:03:15.20 & 18.106 \\
B6-b1 &  18:09:50.480 & -31:40:51.61 & 17.995 \\
B6-b3 &  18:10:19.060 & -31:40:28.19 & 17.705 \\
B6-b4 &  18:10:07.770 & -31:52:41.36 & 17.842 \\
B6-b5 &  18:10:37.380 & -31:40:29.14 & 17.711 \\
B6-b6 &  18:09:49.100 & -31:50:07.66 & 17.793 \\
B6-b8 &  18:09:55.950 & -31:45:46.33 & 18.397 \\
B6-f1 &  18:10:04.460 & -31:41:45.31 & 17.901 \\
B6-f3 &  18:10:17.720 & -31:41:55.20 & 17.255 \\
B6-f5 &  18:10:41.510 & -31:40:11.88 & 17.632 \\
B6-f7 &  18:10:52.300 & -31:46:42.18 & 17.841 \\
B6-f8 &  18:09:56.840 & -31:43:22.56 & 17.263 \\
BL-1  &  18:34:58.510 & -34:33:15.24 & 16.905 \\
BL-3  &  18:35:27.510 & -34:31:59.36 & 16.884 \\
BL-4  &  18:35:21.110 & -34:44:48.22 & 16.451 \\
BL-5  &  18:36:01.010 & -34:31:47.91 & 16.911 \\
BL-7  &  18:35:57.260 & -34:38:04.61 & 16.579 \\
\hline
\end{tabular}
\label{tab:basicdata}
\tablefoot{\\
\tablefootmark{a}{Using the same naming convention as \citet{2007A&A...465..799L} for the B3-BW-B6-BL-stars.}
}
\end{table}

The spectra of the 35 stars in the B3, BW, B6, and BL fields analyzed here, are the same that have been analyzed for O in \citet{2006A&A...457L...1Z}, Na, Mg, Al in \citet{2007A&A...465..799L}, Mn \citet{2013A&A...559A...5B}, and Ba, La, Ce, Nd, Eu in \citet{2016A&A...586A...1V}. These observations were carried out May-Aug 2003-2004. Since the fiber-array FLAMES was used in combination with UVES, seven stars could be observed in each pointing. Four spectra have been excluded from the analysis because the determined stellar parameters are outside of the parameter space tested in Paper I: for one star, B3-b3, we derive a large $\log g$ of 3.23, making it possible to be a foreground disk star, for two stars, B3-b4 and B6-f2, we need to use atmospheric turbulence parameters outside of the ranges $1.0<v_{\mathrm{mic}}<2.0$ and $1.0<v_{\mathrm{mac}}<8.0$  that is making the determination of the surface gravity uncertain, and for one star, BW-f4, we derive a [Fe/H]$=-1.55$, that is lower than any of the stars analyzed in Paper I, and we are not certain how well our stellar parameter determination works in this regime.

The 11 new stars in Sagittarius Window were observed in the same way using the same telescope, instrument, and setting in service mode during Aug 2011 (ESO program 085.B-0552(A)).

The total integration time in each setting was 5-12 hours depending on extinction. The achieved S/N is listed in Table \ref{tab:stellarparams}. The resolving power of the spectra is 47000 and the spectra cover the region 5800~\AA~to 6800~\AA. 
 
% ##################### Analysis ##################
\section{Analysis and results} \label{sec:analysis}
The spectra were analyzed using the exact same techniques and spectral lines used to analyze the Solar neighborhood stars in Paper I. In short, the software \texttt{Spectroscopy Made Easy}, SME \citep{1996A&AS..118..595V} was used to, via $\chi^2$-minimization of a synthetic spectrum and the observed spectrum, determine the stellar parameters as well as the abundances. In the analysis we used spherical symmetric, [$\alpha$/Fe]-enhanced, LTE MARCS-models. Furthermore, NLTE-corrections were used for the iron-lines \citep{2012MNRAS.427...50L}.

\subsection{Reference sample}
\begin{figure*}[htp]
\centering
\includegraphics[width=180mm]{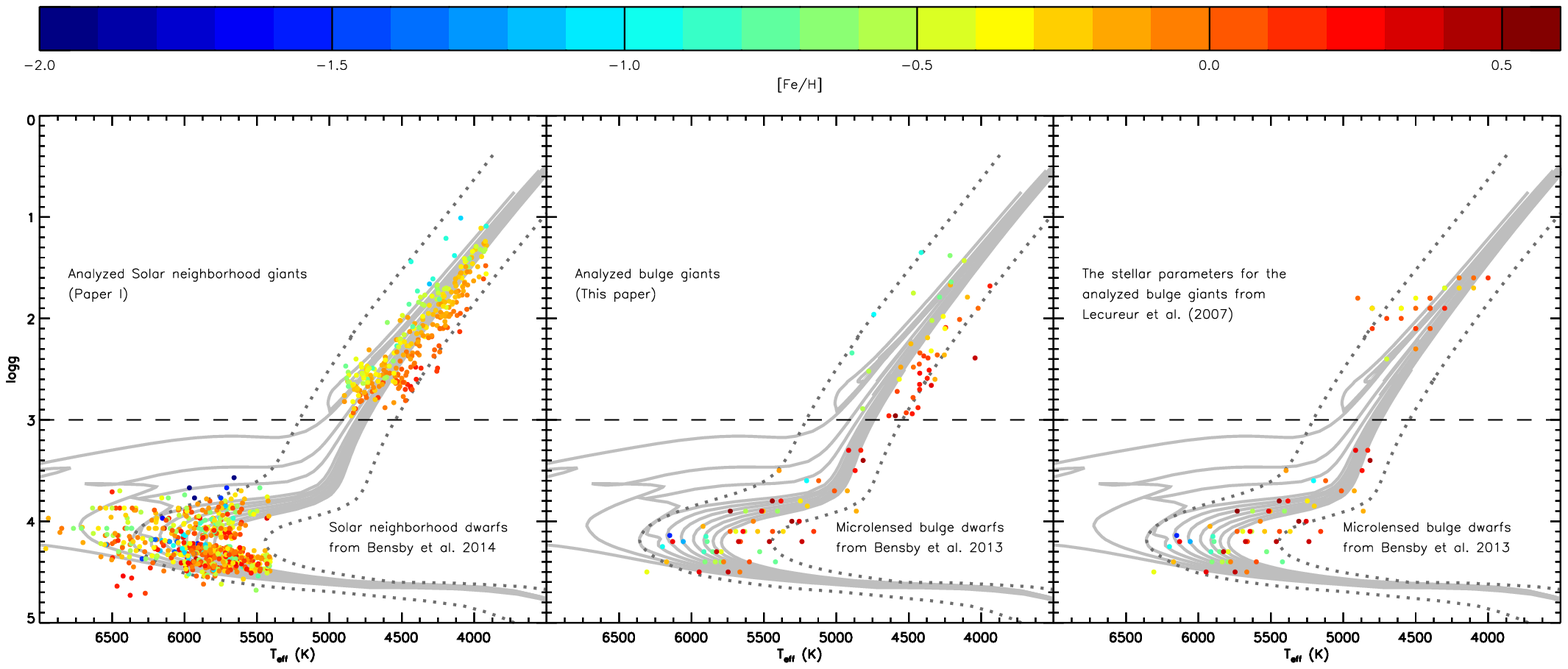}
\caption{ HR-diagrams for the program stars and the reference sample from Paper I. Also shown are the microlensed bulge dwarfs of \citet{2013A&A...549A.147B} and its local disk comparison sample \citep{2014A&A...562A..71B} as well as the stellar parameters for the here analyzed bulge sample as determined by \citet{2007A&A...465..799L}. As a guide for the eye, isochrones with [Fe/H]=0.0 and ages 1-10 Gyr are plotted using solid light gray lines. Furthermore, one isochrone with [Fe/H]=-1.0 and age 10 Gyr, and one with [Fe/H]=+0.5 and age 10 Gyr are plotted using dotted dark grey lines \citep{2012MNRAS.427..127B}.}
\label{fig:hr-dia}
\end{figure*}

The main point of Paper I was to analyze a Solar neighborhood reference sample of giant stars similar to the bulge stars analyzed here. The HR-diagram of this Solar neighborhood reference sample and the Solar neighborhood dwarf stars of \citet{2014A&A...562A..71B} are shown in the leftmost panel of Figure \ref{fig:hr-dia}.

\subsection{The bulge sample}
In order to enable a strictly differential comparison to the reference sample, we have re-determined the stellar parameters as well as the abundances also for the previously published B3-BW-B6-BL-stars using the exact same purely spectroscopic analysis as is used for the Solar neighborhood reference sample of Paper I. The resulting stellar parameters are plotted as a HR-diagram in the middle panel of Figure \ref{fig:hr-dia}. As a comparison, the parameters for the same stars as determined in \citet{2007A&A...465..799L} are shown in the rightmost panel. As can be seen, the largest differences between the two sets of stellar parameters are seen in the surface gravities, where our results are spread out along the red giant branch, while this is not shown in the previous stellar parameters of \citet{2007A&A...465..799L}. Furthermore, our results, in contrast to the older parameters, are sorted in [Fe/H] as expected in Figure \ref{fig:hr-dia}. This, together with the careful evaluation of the method used (in Paper I), gives us confidence in our determined stellar parameters.

\begin{figure*}[htp]
\centering
\includegraphics[width=170mm]{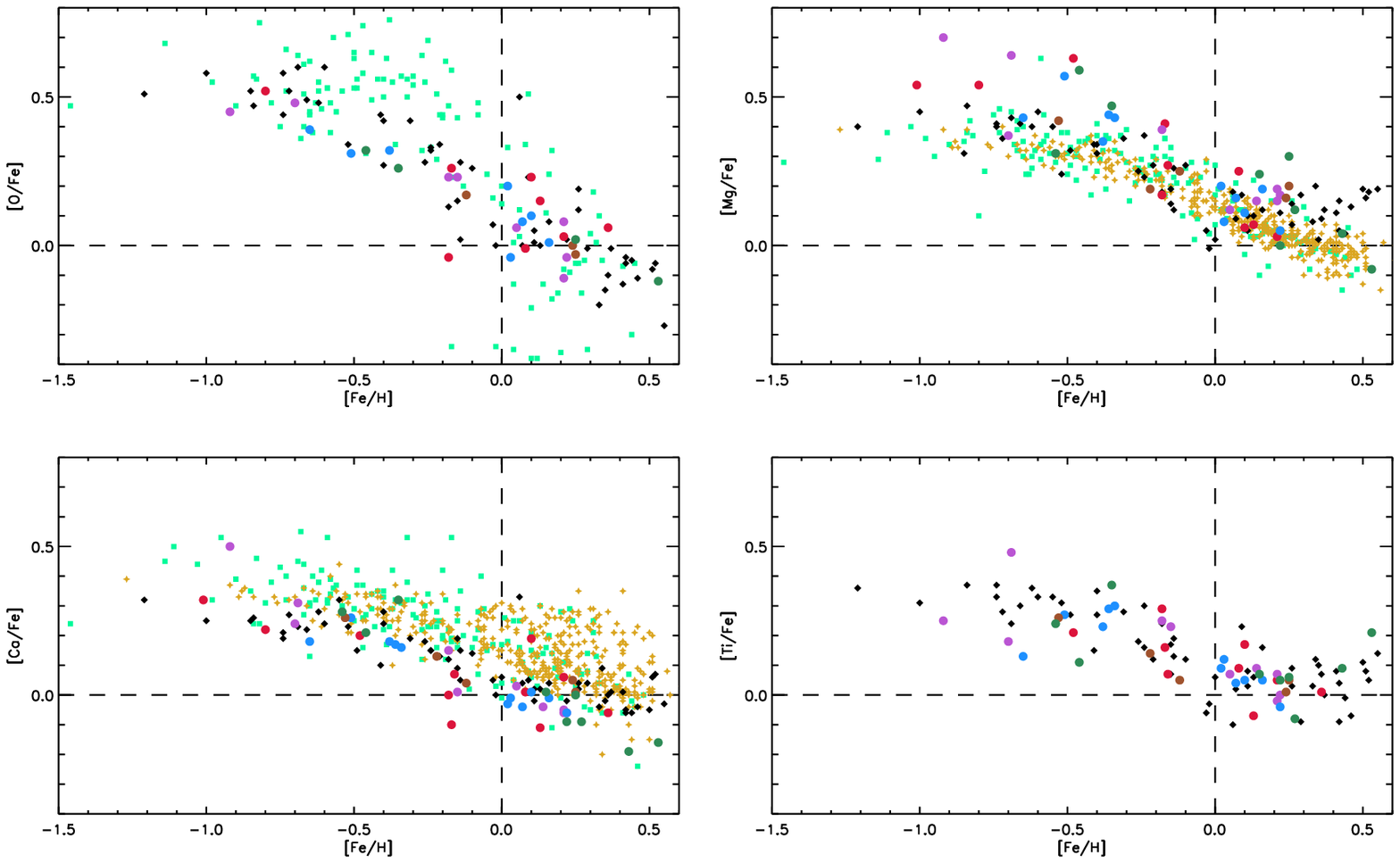}
\caption{[X/Fe] vs. [Fe/H] for the analyzed bulge giants and some previous works. The stars are color-coded as the corresponding fields in Figure~\ref{fig:bulge_fields}, with black diamonds representing the microlensed Bulge dwarfs of \citet{2013A&A...549A.147B}, light green squares showing the abundances of \citet{2014AJ....148...67J}, and golden stars showing the abundances of the high-resolution sample of the GIBS survey \citep{2015A&A...584A..46G}.  We use A(O)$_{\odot}=8.69$, A(Mg)$_{\odot}=7.60$, A(Ca)$_{\odot}=6.34$, A(Ti)$_{\odot}=4.95$, and A(Fe)$_{\odot}=7.50$ \citep{2009ARA&A..47..481A}.}
\label{fig:all_abundances}
\end{figure*}

The determined parameters and abundances are listed in Table \ref{tab:stellarparams}. Furthermore, the determined abundances are shown in Figure \ref{fig:all_abundances} together with the microlensed bulge dwarfs of \citet{2013A&A...549A.147B}, the abundances of \citet{2014AJ....148...67J}, and the GIBS survey \citep{2015A&A...584A..46G}.

\begin{table*}[htp]
\caption{Determined stellar parameters and abundances for the observed bulge giants.}
\begin{tabular}{l c c r c c c c c c }
\hline
\hline
Star &  $T_\textrm{eff}$ & $\log g$ & [Fe/H]\tablefootmark{a} & $\xi_\textrm{micro}$ & S/N\tablefootmark{b} & $\log \epsilon(\mathrm{O})$ & $\log \epsilon(\mathrm{Mg})$ &  $\log \epsilon(\mathrm{Ca})$ & $\log \epsilon(\mathrm{Ti})$\\
     & [K]               & (cgs)    &                   &      [km\,s$^{-1}$]       \\
\hline
SW-09 &   4095 &   1.79 &  -0.18 &   1.32 &    16 &   8.47 &   7.59 &   6.16 &   5.06 \\
SW-15 &   4741 &   1.96 &  -1.01 &   1.62 &    15 &    ... &   7.13 &   5.65 &    ... \\
SW-17 &   4245 &   2.09 &   0.21 &   1.44 &    11 &   8.93 &   7.84 &   6.61 &   5.21 \\
SW-18 &   4212 &   1.67 &  -0.16 &   1.49 &    14 &    ... &   7.71 &   6.25 &   4.86 \\
SW-27 &   4423 &   2.34 &   0.08 &   1.60 &    13 &   8.76 &   7.93 &   6.43 &   5.12 \\
SW-28 &   4254 &   2.36 &  -0.17 &   1.44 &    16 &   8.78 &   7.84 &   6.07 &   4.94 \\
SW-33 &   4580 &   2.72 &   0.13 &   1.39 &    14 &   8.97 &   7.80 &   6.36 &   5.01 \\
SW-34 &   4468 &   1.75 &  -0.48 &   1.63 &    12 &    ... &   7.75 &   6.06 &   4.68 \\
SW-43 &   4892 &   2.34 &  -0.80 &   1.84 &    16 &   8.41 &   7.34 &   5.76 &    ... \\
SW-71 &   4344 &   2.66 &   0.36 &   1.31 &    14 &   9.11 &    ... &   6.64 &   5.32 \\
SW-76 &   4427 &   2.45 &   0.10 &   2.00 &    12 &   9.02 &   7.76 &   6.63 &   5.22 \\
\hline
      B3-b1 &   4414 &   1.35 &  -0.92 &   1.41 &    21 &   8.22 &   7.38 &   5.92 &   4.28 \\
      B3-b5 &   4425 &   2.70 &   0.22 &   1.43 &    43 &   8.87 &   7.99 &   6.50 &   5.17 \\
      B3-b7 &   4303 &   2.36 &   0.05 &   1.58 &    38 &   8.80 &   7.77 &   6.42 &   5.07 \\
      B3-b8 &   4287 &   1.79 &  -0.70 &   1.46 &    65 &   8.47 &   7.27 &   5.88 &   4.43 \\
      B3-f1 &   4485 &   2.25 &  -0.18 &   1.88 &    31 &   8.74 &   7.81 &   6.31 &   5.02 \\
      B3-f2 &   4207 &   1.64 &  -0.69 &   1.74 &    22 &    ... &   7.55 &   5.96 &   4.74 \\
      B3-f3 &   4637 &   2.96 &   0.21 &   1.89 &    31 &   8.98 &   8.00 &   6.49 &   5.14 \\
      B3-f4 &   4319 &   2.60 &  -0.15 &   1.50 &    11 &   8.77 &    ... &   6.20 &   5.03 \\
      B3-f7 &   4517 &   2.93 &   0.14 &   1.55 &    24 &    ... &   7.89 &   6.44 &   5.18 \\
      B3-f8 &   4436 &   2.88 &   0.21 &   1.54 &    63 &   8.79 &   7.96 &   6.50 &   5.23 \\
      BW-b1 &   4042 &   2.39 &   0.43 &   1.43 &    29 &    ... &   8.07 &   6.58 &   5.47 \\
      BW-b2 &   4367 &   2.39 &   0.15 &   1.68 &    20 &    ... &   7.99 &   6.50 &   5.17 \\
      BW-b5 &   3939 &   1.68 &   0.22 &   1.31 &    43 &    ... &   7.82 &   6.47 &   5.22 \\
      BW-b6 &   4262 &   1.98 &  -0.35 &   1.44 &    23 &   8.60 &   7.72 &   6.31 &   4.97 \\
      BW-b8 &   4424 &   2.54 &   0.27 &   1.52 &    44 &    ... &   7.99 &   6.52 &   5.14 \\
      BW-f1 &   4359 &   2.51 &   0.25 &   1.93 &    37 &   8.96 &   8.15 &   6.59 &   5.26 \\
      BW-f5 &   4818 &   2.89 &  -0.54 &   1.29 &    39 &    ... &   7.37 &   6.08 &   4.65 \\
      BW-f6 &   4117 &   1.43 &  -0.46 &   1.69 &    33 &   8.55 &   7.73 &   6.09 &   4.60 \\
      BW-f7 &   4592 &   2.96 &   0.53 &   1.50 &    15 &   9.10 &   8.05 &   6.71 &   5.69 \\
      B6-b1 &   4372 &   2.59 &   0.22 &   1.57 &    51 &    ... &   7.87 &   6.50 &   5.13 \\
      B6-b3 &   4468 &   2.48 &   0.02 &   1.67 &    59 &   8.91 &   7.82 &   6.33 &   5.06 \\
      B6-b4 &   4215 &   1.38 &  -0.65 &   1.68 &    41 &   8.43 &   7.38 &   5.87 &   4.43 \\
      B6-b5 &   4340 &   2.02 &  -0.51 &   1.34 &    54 &   8.49 &   7.66 &   6.09 &   4.71 \\
      B6-b6 &   4396 &   2.37 &   0.16 &   1.77 &    44 &   8.86 &   7.95 &   6.49 &   5.16 \\
      B6-b8 &   4021 &   1.90 &   0.03 &   1.45 &    55 &   8.68 &   7.71 &   6.36 &   5.10 \\
      B6-f1 &   4149 &   2.01 &   0.07 &   1.65 &    77 &   8.84 &   7.83 &   6.37 &   5.06 \\
      B6-f3 &   4565 &   2.60 &  -0.38 &   1.28 &    82 &   8.63 &   7.57 &   6.14 &   4.80 \\
      B6-f5 &   4345 &   2.32 &  -0.36 &   1.41 &    32 &    ... &   7.68 &   6.15 &   4.88 \\
      B6-f7 &   4250 &   2.10 &  -0.34 &   1.65 &    29 &    ... &   7.69 &   6.16 &   4.91 \\
      B6-f8 &   4470 &   2.78 &   0.10 &   1.30 &    81 &   8.89 &   7.81 &   6.45 &   5.10 \\
       BL-1 &   4370 &   2.19 &  -0.22 &   1.50 &    38 &    ... &   7.57 &   6.25 &   4.87 \\
       BL-3 &   4555 &   2.48 &  -0.12 &   1.53 &    57 &   8.74 &   7.73 &   6.26 &   4.88 \\
       BL-4 &   4476 &   2.94 &   0.24 &   1.41 &    36 &   8.93 &   8.00 &   6.63 &   5.20 \\
       BL-5 &   4425 &   2.65 &   0.25 &   1.68 &    58 &   8.91 &   8.05 &   6.60 &   5.25 \\
       BL-7 &   4776 &   2.52 &  -0.53 &   1.53 &    60 &    ... &   7.49 &   6.07 &   4.68 \\
\hline
\end{tabular}
\label{tab:stellarparams}
\tablefoot{\\
\tablefootmark{a}{ We use A(Fe)$_{\odot}=7.50$ \citep{2009ARA&A..47..481A}.\\}
\tablefootmark{b}{S/N per datapoint as measured by the IDL-routine \texttt{der\textunderscore snr.pro}, see \href{http://www.stecf.org/software/ASTROsoft/DER\textunderscore SNR}{http://www.stecf.org/software/ASTROsoft/DER\textunderscore SNR}\\}
}
\end{table*}

% ##################### Discussion ##################
\section{Discussion} \label{sec:discussion}
We are not able to see any trends in metallicity nor abundances across the different fields, why we in the following handle the entire sample as a bulge sample. To see possible trends more stars in every field are needed.

It is hard to estimate the age of giant stars, but from the isochrones in Figure \ref{fig:hr-dia}, one can see a slight splitting up with respect of age for the giants with highest gravities (close to $\log g=3$). This might possibly be visible in a slight split/spread of the solar metallicity (orange in the plot) stars of the solar neighborhood sample, while the same effect is not clearly visible in the bulge sample, which is expected since the bulge stars are predominately old \citep[e.g.,][]{2008ApJ...684.1110C}, \citep[but at the same time, see][]{2013A&A...549A.147B}. From Figure \ref{fig:hr-dia} it is also obvious that the bulge stars generally are more metal-rich than the giants found in the Solar neighborhood.

\subsection{Comparison to other studies}
From Figure \ref{fig:all_abundances}, one can see that the trends of our stars and the microlensed dwarfs \citep{2013A&A...549A.147B} are quite similar, and the scatter seem to be rather similar, with our stars possibly showing marginally higher scatter.
Our [O/Fe] vs. [Fe/H] trend is much less scattered and less steep than that of \citet{2014AJ....148...67J}. The differences are likely due to the large uncertainties inherent in determining the oxygen abundance from the 6300~Å [\ion{O}{i}]-line in the relatively low resolution spectra of \citet{2014AJ....148...67J}. For example, \citet{2011A&A...530A..54G} avoid determining the oxygen abundance all-together from the exact same data due to these uncertainties.
When it comes to our [Ca/Fe] vs. [Fe/H] trend, it is much tighter and less alpha enhanced than the corresponding trends of \citet{2014AJ....148...67J} and \citet{2015A&A...584A..46G}. These differences may be attributed to our stellar parameters being more accurate due to our larger wavelength coverage, higher resolution, and thorough tests of our method in Paper I.
On the other hand, the [Mg/Fe] vs. [Fe/H] trends of \citet{2014AJ....148...67J} and \citet{2015A&A...584A..46G} are less alpha enhanced and tighter than ours. The reason for our data showing a larger scatter is not clear: all three works use the same three \ion{Mg}{i} lines around 6318-19 Å and our data has higher resolution suggesting that our data, at least theoretically, should be of higher quality. However, these three lines have several difficulties: first of all, they have uncertain $gf$-values. \citet{2014AJ....148...67J} and \citet{2015A&A...584A..46G} use astrophysical values and we use the (very similar) results of \citet{2016arXiv161107000P}. Secondly, the lines are affected by an autoionizing \ion{Ca}{i}-line producing a very wide depression of the spectrum. \citet{2014AJ....148...67J}, like us, solve this problem by setting a local pseudo-continuum around the \ion{Mg}{i} lines, while \citet{2015A&A...584A..46G} both model the autoionizing \ion{Ca}{i}-line using their determined Ca-abundance, \emph{and} place a local continuum to get rid of possible residual mis-matches between the observed and synthetic spectra. Thirdly, the lines are in a region affected by telluric lines. \citet{2014AJ....148...67J} remove these by division by an observed `telluric' spectrum, while we and \citet{2015A&A...584A..46G} simply avoid using the Mg-lines visibly affected by telluric contamination. To conclude, a possible explanation for our more scattered [Mg/Fe] vs. [Fe/H] trend might be that the (necessary) lower S/N of our higher resolution data makes the continuum-placement more difficult, and our tendency to derive higher [Mg/Fe] might be due to the lower S/N making it harder to identify and avoid telluric lines, implying that we would derive too high magnesium abundances in the cases where we possibly fail to identify a telluric line.

Several previous studies, but not all, have found different trends in [O/Fe] and [Mg/Fe] in the bulge, which is often attributed to a higher degree massive stars in the bulge compared to the Solar neighborhood. One example is \citet{2007A&A...465..799L}, who used 35 of the same spectra as we do,  opening up for an interesting comparison. Therefore, we have plotted the oxygen abundances of \citet{2006A&A...457L...1Z} (these are the very same abundances also presented in \citet{2007A&A...465..799L}), and the magnesium abundances of \citet{2007A&A...465..799L} in Figure \ref{fig:sn-bulge_abundances}. Our re-analysis show a similar, but slightly less scattered, oxygen-trend as \citet{2006A&A...457L...1Z}, while our magnesium-trend is lower in [Mg/Fe] for [Fe/H]$>-0.5$, showing a rather thick-disk-like trend at odds with what is found in \citet{2007A&A...465..799L}. We believe that most of these differences can be attributed to our new all-spectroscopic stellar parameters, but also to the different handling of the autoionizing \ion{Ca}{i}-line affecting the derived magnesium abundances: \citet{2007A&A...465..799L} model this line to get rid of its influence in spite of its uncertain spectroscopic data, while we avoid synthesizing it and instead place a local pseudo-continuum around the three \ion{Mg}{i}-lines (for some example of their modeling of this line, see Figures 3 and 5 in \citet{2007A&A...465..799L}).

There are several ongoing large spectroscopic projects surveying the entire Galaxy, and including the bulge. For example the APOGEE survey \citep{2011AJ....142...72E} has observed a wealth of stars (over 150~000), with several fields towards the bulge. APOGEE has the advantage of observing in the H-band, reducing the problem with extinction of light due to dust, but the rather small diameter of the telescope used, means that the stars that actually are \emph{in} the bulge are the most luminous giants, that are the hardest to analyze. As of yet there has not been any APOGEE-paper on the bulge, but only on the very special stellar population of the absolute galactic centre \citep{2015A&A...584A..45S}. 

The Gaia ESO-survey has some fields in the bulge, but has sofar not published any comparison between the alpha elemental trends of the local disk and the bulge, but only an investigation on the metallicity and kinematic trends of the bulge \citep{2014A&A...569A.103R}.

\subsection{Comparison to the Solar neighborhood sample}
As has been mentioned several times before, the abundance trends found in the bulge must be compared to similarly determined trends in the disk: most importantly the type of stars and the spectral lines used in the analysis should be the same to minimize systematic differences. Ideally also the quality of the spectra should be the same - the resolution and S/N - but this is harder to obtain: it is impossible to obtain the same S/N for the faint bulge giants as for the bright nearby disk giants, and for this difference in magnitude, possibly the same telescope/spectrometer cannot be used in both cases. In our case we used FIES \citep{2014AN....335...41T} at NOT and data retrieved from the NARVAL and ESPaDOnS spectral archive PolarBase \citep{2014PASP..126..469P} to collect spectra for the Solar neighborhood sample of Paper~I, while we used UVES/FLAMES at VLT for the bulge sample. The FIES and PolarBase spectra have a resolving power of 67000 and 65000, respectively, and high S/N (typically around 100), while the UVES/FLAMES spectra have $R \sim 47000$ and much lower S/N (see Table \ref{tab:stellarparams}). The effect of this difference in spectral quality is expected to manifest itself as more scatter in the bulge trends.

\begin{figure*}[htp]
\centering
\includegraphics[width=170mm]{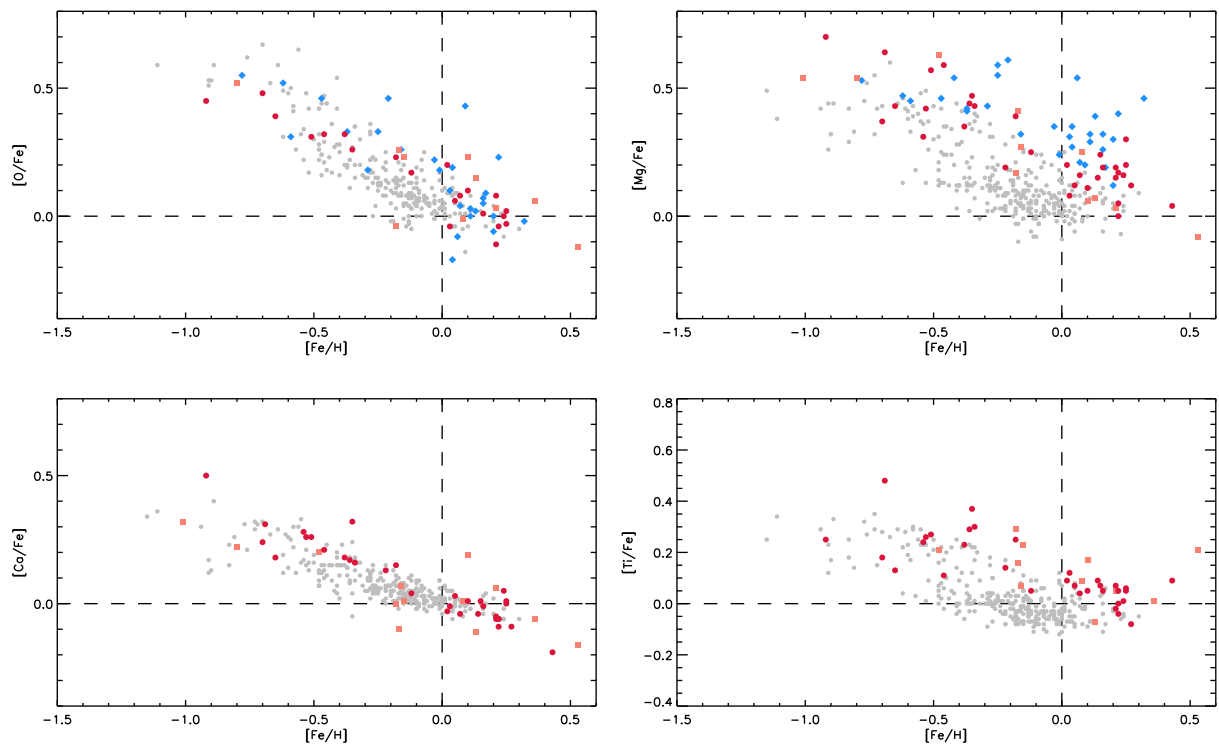}
\caption{[X/Fe] vs. [Fe/H]. From our investigation of the impact of S/N on the abundances (Paper I, Figure 2), we conclude that the uncertainties of the abundances are multiplying for S/N<20. Therefore, we plot our results for the bulge spectra with S/N>20 using red dots and the results from those with S/N<20 using pink squares. The Solar neighborhood reference sample of giants from Paper I is plotted in gray. Previous abundances for the exact same spectra from \citet{2006A&A...457L...1Z} (oxygen) and \citet{2007A&A...465..799L} (magnesium) are plotted using blue diamonds.}
\label{fig:sn-bulge_abundances}
\end{figure*}

In Figure \ref{fig:sn-bulge_abundances}, where we compare the abundance trends from our Solar neighborhood sample of Paper I to that of the bulge stars of this article, it is obvious that the abundance trends in the bulge are indeed not as tight as the trends from the Solar neighborhood. Since the type of stars analyzed, and the lines used are the same, this larger spread can only be attributed to lower S/N in the bulge-observations. Looking at Table \ref{tab:stellarparams}, and comparing to Figure 2 in Paper I where we investigate the impact of S/N on the stellar parameters and the abundances, all SW-stars are expected to have an uncertainty in the [X/Fe] abundance ratio of around 0.2 dex (standard deviation) stemming from the S/N alone. The B3-BW-B6-BL-stars generally have higher S/N and are expected to show lower uncertainties due to the S/N, in general around 0.1 dex (standard deviation) for the [X/Fe] abundance ratios. As mentioned earlier, the scatter in the [Mg/Fe]-trend for the bulge stars seems higher than for the other elements, and is possibly   slightly enhanced compared to the thick disk, which is not seen for the other elements. This strongly suggests that the larger scatter in [Mg/Fe] for the bulge stars is linked to both the low S/N making it hard to place the continuum and identify telluric lines in the spectrum.

Comparing the abundance trends of Figure \ref{fig:sn-bulge_abundances}, we find the bulge trends to generally follow that of the local thick disk, but possibly tracing the upper envelope in the case of magnesium, calcium and titanium, while the bulge oxygen trend more seem to follow the lower envelope of the local thick disk (or upper envelope of the local thin disk). These lower oxygen abundances could potentially be explained by the lower S/N of the bulge stars: from Figure 2 in Paper I, an asymmetry for the lowest S/N is seen in the oxygen abundance, suggesting that lower oxygen abundances are derived for lower S/N. In general, the oxygen abundance is expected to be more sensitive to lower quality of the spectra, since it is based on a single line, in contrast to the magnesium, calcium, and titanium abundances. On the other hand, the oxygen and calcium abundance trends are the tightest of the four, suggesting that the determined surface gravity is precise: the oxygen and calcium abundances are mainly dependent on the surface gravity, as is shown in Table 3 in Paper I for oxygen, and is evident in the case of calcium since it is used to constrain the surface gravity.

\begin{figure*}[htp]
\centering
\includegraphics[width=170mm]{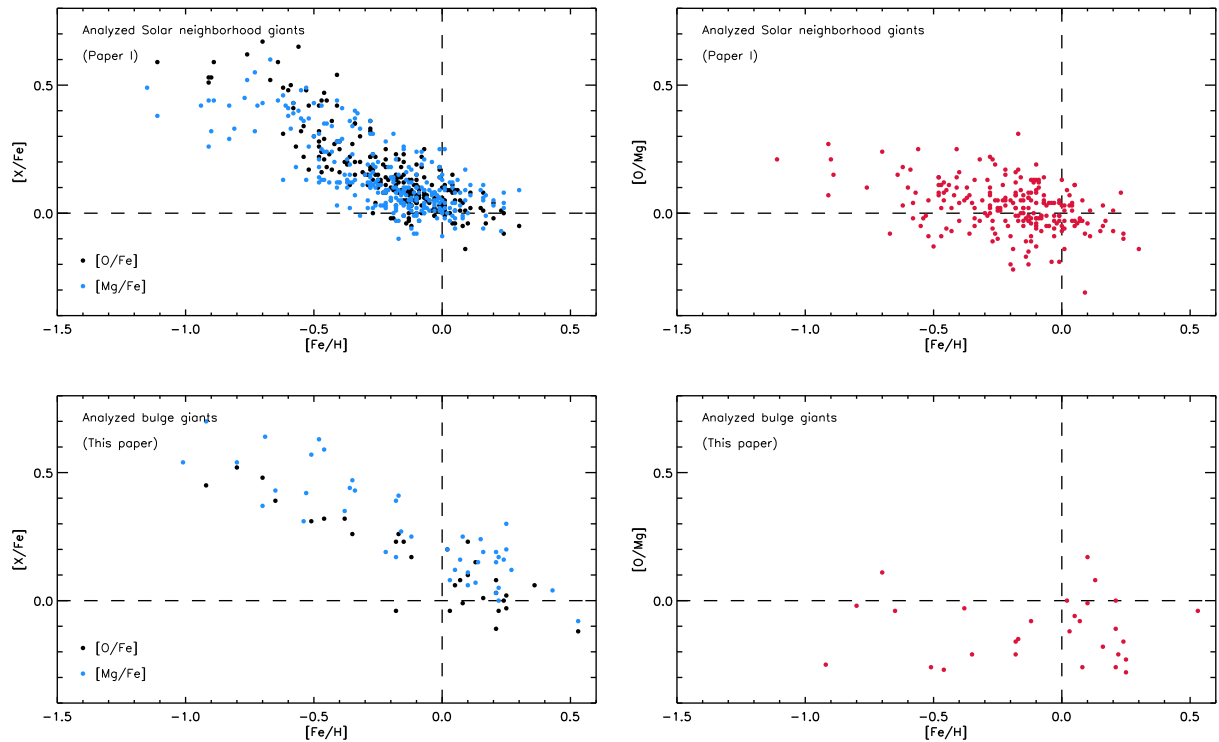}
\caption{The plots to the left show [O/Fe] and [Mg/Fe] vs. [Fe/H] for the Solar neighborhood and bulge giants, respectively. The rightmost plots show [O/Mg] vs. [Fe/H] for the Solar neighborhood and bulge giants, respectively. }
\label{fig:omg}
\end{figure*}

In spite of the slight differences of oxygen and magnesium when comparing our bulge sample to our local disk sample in Figure \ref{fig:sn-bulge_abundances}, we cannot see any evidence for [O/Fe] and [Mg/Fe] showing different trends in our bulge data, see Figure \ref{fig:omg}. The [O/Mg] in our local disk sample is around zero for all metallicities, while the [O/Mg] in our bulge data is negative, but still constant. This difference might be on account of our oxygen abundances in the bulge possibly being systematically too low on account of the lower S/N in the bulge spectra. Furthermore, a slightly higher magnesium abundances in the bulge trend can possibly be attributed to difficulties in avoiding telluric lines in the more noisy bulge spectra as mentioned earlier.  

From Figure \ref{fig:sn-bulge_abundances}, we see no clear evidence for different positions of the knees of the bulge population and the thick disk population, thereby corroborating \citet{2015A&A...584A..46G}, but the conclusion is weak. To make a firmer statement, we would ideally need a larger bulge sample extending to lower metallicities, and more thick disk stars in our local sample.

To resolve the question about the possible higher knee in the alpha elemental abundance plots for the bulge as compared to the local thick disk, one would need two decent sized stellar samples: one from the bulge and one from the local disk. This can be reached in several ways, some of which are listed below: 
\begin{itemize}
\item The investigation of \citet{2013A&A...549A.147B} has a large and representative sample of 714 local dwarf stars, while the bulge sample is much smaller with 58 microlensed dwarfs. Ideally, the bulge sample should be enlarged, but with the unpredictability of the microlensing events, this is not readily done. Also, the type of stars in the two samples are not exactly the same with the local sample being F and G dwarf stars, while the bulge sample contains several slightly cooler subgiant stars, see Figure \ref{fig:hr-dia}.
\item The investigation of \citet{2014AJ....148...67J} and \citet{2015A&A...584A..46G}, on the other hand, both have large bulge sample of hundreds of giant stars, with a very tight and un-scattered [Mg/Fe]-trend, see Figure \ref{fig:all_abundances}. However, they both have a small similarly observed and analyzed local disk sample of giants to contrast their bulge-trend to. Furthermore, their other $\alpha$-elements show larger scatter than our trends.
\item Our investigation has a local disk sample of 291 giants, but would benefit from having more thick-disk stars. The bulge-sample consists of 46 giants of very similar types to the local sample, see Figure \ref{fig:hr-dia}.
\end{itemize}

The strategy used here, observing K-giants in the optical with high resolution spectroscopy, has the upside that it is easy to find and observe suitable, very similar local disk stars and also telescopes/instruments with which to carry out these observations. The downside is the long integration times needed for the bulge observations. However, FLAMES/UVES offers the ability to observe seven stars simultaneously, resulting in about one hour telescope-time per star, similar to the amount of time spent per microlensed dwarf star in \citet{2013A&A...549A.147B}.

For the future, a similar methodology as presented here but performed in the near-IR H and/or K bands would be rewarding. This is indeed possible now with the new cross-dispersed high-resolution, near-IR spectrometers recently available or planned for \citep[see e.g.][]{2014SPIE.9147E..1DP,2014SPIE.9147E..1EO}, but to do so, a serious effort in exploring usable and reliable spectral features in the near-IR needs to be adressed.

\section{Conclusions} \label{sec:conclusion}
We have determined the abundances of oxygen, magnesium, calcium, and titanium in a sample of 46 bulge K-giants, 35 of which have been analyzed for oxygen and magnesium in previous works \citep{2006A&A...457L...1Z,2007A&A...465..799L}, and compare the abundances to those of 291 similarly analyzed K-giants in the solar neighborhood.

To conclude, our re-analysis of the bulge oxygen abundances from \citet{2006A&A...457L...1Z} and the magnesium abundances from \citet{2007A&A...465..799L}, result in similar oxygen trends, while we do not see the high [Mg/Fe]-values for the highest [Fe/H]-stars. Thereby we contradict  \citet{2007A&A...465..799L} and their claim that the oxygen and magnesium trends are very different in the bulge.

Furthermore, the question of a possible shift in position of the knee in the [$\alpha$/Fe]-plot in the bulge as compared to the local disk is not unambiguously answered.

\begin{acknowledgements}
This research has been partly supported by the Royal Physiographic Society in Lund, Stiftelsen Walter Gyllenbergs fond. N.R. acknowledges support from the Swedish Research Council, VR (project number 2014-5640). M.Z. acknowledges support by the Ministry of Economy, Development, and Tourism's Millennium Science Initiative through grant IC120009, awarded to The Millennium Institute of Astrophysics (MAS), by  Fondecyt Regular 1150345 and by the BASAL-CATA Center for Astrophysics and Associated Technologies PFB-06. This publication made use of the SIMBAD database, operated at CDS, Strasbourg, France, NASA's Astrophysics Data System, and the VALD database, operated at Uppsala University, the Institute of Astronomy RAS in Moscow, and the University of Vienna.
\end{acknowledgements}

\bibliography{/Users/henrik/Documents/Bibliografi/papers.bib,/Users/henrik/Documents/Bibliografi/kurucz.bib,/Users/henrik/Documents/Bibliografi/sulphur.bib,/Users/henrik/Documents/Bibliografi/GESreferencesv5all.bib}
\bibliographystyle{aa}

\end{document}